\documentstyle[12pt]{article}

\newcommand{\eq}[1]{eq.(\ref{#1})}

\def\be{\begin{equation}}
\def\ee{\end{equation}}

\begin{document}
\begin{flushright}
hep-ph/9212275 \\
UCLA/92/TEP/49 \\
December 1992\\
\end{flushright}
\vspace{0.5in}
\begin{center}
{\bf\large Light fermions in composite models} \\
\vspace{0.4in}
S. Yu. Khlebnikov\footnote{On leave of absence
from Institute for Nuclear Research of the Academy
of Sciences, Moscow 117312 Russia.} and R. D. Peccei\\
\vspace{0.2in}
{\it Department of Physics, University of California,
Los Angeles, CA 90024, USA} \\
\vspace{0.7in}
{\bf Abstract} \\
\end{center}
In preon models based on chiral gauge theories, we show that light
composite fermions can ensue as a result of gauging a subset of preons
in a vector-like manner. After demonstrating how this mechanism works
in a toy example, we construct a one generation model of quarks which
admits a hierarchy between the up and down quark masses as well as
between these masses and the compositeness scale.
In simple extensions of this model to more generations we discuss
the challenges of obtaining any quark mixing. Some possible
phenomenological implications of scenarios where quarks and leptons
which are heavier are also less pointlike are also considered.

\newpage
\section{Introduction}
The idea that quarks and leptons are composite (for a review, see
ref.\cite{Peccei2}) has been pursued as an explanation
for the observed fermionic masses and mixings. There are many problems,
however, met in implementing this idea. One of the main problems is to
understand why quarks and leptons are so light, compared to the inverse
of their 'size'  set by the compositeness scale.
In vector-like gauge theories like QCD, the masses of the composite states
are of the same order as the compositeness
scale. Because of this, QCD-like theories appear ill suited for building
sensible composite models.
It has been suggested \cite{tHooft} that perhaps in nature the masses
of quarks and leptons are protected from being of the order of the
compositeness scale by some approximate chiral symmetry. Again,
this idea argues against vector-like theories, for we know that
in these theories all global chiral symmetries are spontaneously broken.

In view of the above considerations, it appears that chiral gauge theories
are much more natural theories to consider for preon models of
quarks and leptons. With chiral gauge theories, however,
one typically runs into a converse problem. Namely,
unbroken chiral symmetries lead to {\em massless} fermions and it is
difficult to break these symmetries just slightly, so as to make
the resulting bound state fermion masses small but non-zero.
In the literature models have been proposed
which break the chiral symmetries of the preon theory by
explicit mass terms and/or four-fermion
interactions introduced at the fundamental level.
Although some of these models are interesting, introducing such
seed breaking undermines to a great extent the original motivation to make
quarks and leptons composite.

One may arrive at the idea of compositeness from a completely different
perspective.
An attractive, and well-known, way of breaking the electroweak
symmetry dynamically is provided by technicolor interactions
\cite{tc}. However, technicolor interactions themselves cannot generate
fermion masses and one is forced to introduce yet further interactions,
extended technicolor (ETC), to accomplish this task \cite{etc}.
If the known fermions and some technifermions were to be composite,
the preonic theory would invariably produce some effective four-fermion
interactions among these states. These interactions could hopefully
serve as the seeds for fermion mass generation, without the need of
having to introduce new ETC forces (for early work in this direction
see, for example, ref.\cite{early}).
Thus, the idea of compositeness nicely combines with that of technicolor.
It is, however, not clear a priori whether compositeness
can cure some of the familiar technicolor diseases \cite{diseases},
such as having light pseudoGoldstone bosons in the physical spectrum
and sizable flavor-changing neutral currents (FCNC).

The objective of this paper is two-fold. First, we want to show that
in composite models based on chiral gauge theories, it is possible
in principle to generate {\em light} but not massless fermionic
bound states. This mass generation does not necessitate bare preonic
masses (which are, in fact, forbidden!) or non-renormalizable
interactions introduced by hand, but it is entirely dynamical coming
as a result of additional vector-like interactions acting on a subset
of the preons. After demonstrating how this mechanism works in a simple
context we broach the second objective of this paper, which is to study
whether the observed mass pattern of quarks and leptons can be accounted
for by such a scenario, perhaps by incorporating as well some version of
the technicolor idea. Although we have not been totally successful in
our second goal, the semirealistic model which we construct suggests
interesting generic features which may have important phenomenological
consequences.

The most challenging point in trying to construct a realistic model
of this type is related to the issue of quark mixing. Because
in our model preons corresponding to different generations carry different
quantum numbers, it is not possible to introduce a
Cabibbo-Kobayashi-Maskawa (CKM) matrix directly at the preonic
level. Thus, the CKM matrix must ensue as a low-energy phenomenon
and be in principle calculable in terms of the fundamental parameters of
the theory.
As will be seen, it is relatively easy to introduce a hierarchy of masses
for the up- and down-type quarks. However, it is difficult to actually
break all the vestiges of residual family symmetries in the model
considered, so as to actually generate a CKM matrix. Nevertheless, if
a non-trivial CKM matrix were to ensue, it is very natural in these
scenarios that the concomitant FCNC would dominantly affect the heavy
quark sector. Since FCNC effects involving heavy quarks are not thoroughly
studied experimentally, these considerations suggest that the banishing
of all FCNC may not necessarily be the most sensible strategy to adopt
in model building. In this respect,
our philosophy differs from that of recent attempts
\cite{Georgi} to incorporate
FCNC suppressing mechanisms in composite and non-composite
technicolor models.

In the analysis of composite models based on chiral
gauge theories, which we will present, there are several dynamical
assumptions involved. First, it will turn out that in the model
including three
generations (in which mixing, unfortunately, is difficult to obtain),
because of the plethora of fermionic species, the weak, color
and technicolor interactions are not asymptotically free.
Thus, although the weak and strong couplings behave exactly
as in the standard model at low energy, they start to grow above
energies of around 1 TeV when technifermions become relevant.
Asymptotically non-free non-abelian gauge theories have been invoked
previously \cite{nonfree} in the context of extended technicolor
in attempts to alter the naive relation between the ETC scale
and fermionic masses and thus suppress FCNC effects. We have
nothing to add here as far as the dynamics of such theories is
concerned, nor do we rely on this FCNC suppression in our
further discussion.
We merely will assume that such theories can be made consistent,
at least in the presence of a cutoff, and that for vector-like
theories chirality is spontaneously broken, essentially in the
same way as it happens in ordinary QCD.
We note, however, that in the one-generation model described
in Sect.3 color and technicolor are asymptotically free.

For the analysis of the vector-like pieces of our models a
second set of dynamical assumptions enters.
If all the preons were massive we could use mass inequalities
\cite{inequalities} to argue that the vectorial global symmetries
are preserved by vector-like gauge theories, while chiral global
symmetries are spontaneously broken. It then would follow that
in the limiting case when the bare masses are taken to zero, the vacuum
with unbroken vectorial symmetries and broken chiral ones either
remains the true ground state or is degenerate with it (see Vafa and
Witten in ref.\cite{inequalities}). Even though in our models we cannot
contemplate taking this limit, we shall assume that the former applies,
thus neglecting the possibility of accidental
degeneracy. Note that the mass inequalities hold irrespectively of
whether the vector-like gauge theory is asymptotically free or not.
In sect.4 we will also discuss the relevance of departures of technicolor
theory from vector-like behavior due to additional
interactions, remnants from the preonic theory. It is
interesting to understand if such interactions may lead to the breakdown
of vectorial symmetries, in particular those associated with
family numbers.

Finally, in the analysis of the chiral gauge components of our
models, we rely heavily on the complementarity principle
\cite{complementarity} to ascertain the pattern of symmetry
breakdown. Because the idea of complementarity may not be as well
known, we will describe it briefly in the following section.

The outline of the paper is as follows. In sect.2 we describe
how, by gauging vector-like subgroups, one can actually generate
light fermions in a chiral gauge preon theory, illustrating the
mechanism with a simple but unrealistic model.
In sect.3 we describe a semirealistic, one generation, composite model
of quarks which can easily encompass a mass hierarchy.
Generations, the issue of quark mixing and the concomitant appearance
of FCNC in such models are broached in Sect.4. Here, even though
no realistic models are actually constructed, some of the possible
phenomenological implications of this type of scenarios are noted.
Finally, Sect.5 contains our conclusions.

\section{Mass generation in a chiral gauge theory}
The chiral gauge component of the theory discussed in this section
was studied long ago by Bars and Yankielowicz \cite{BY}
and is described in some
detail in connection with the problem of mass
generation in ref.\cite{Peccei2}.
The model is based on the gauge group $SU_{gauge}(N)$ and has
$N+4$ copies of massless chiral fermions $F_{ia}$
($i=1,...,N$, ~$a=1,...,N+4$) in the fundamental representation
and a single copy $S^{ij}$ in the conjugate symmetric representation.
This content
is free from gauge anomalies. The model may be analyzed by using the
complementarity principle \cite{complementarity}.
The most attractive channel \cite{MAC}
(that is the one with the largest relative Casimir operator) favors a
condensate
\be
\langle F_{ia} S^{ij} \rangle \equiv \langle \Phi^j_a \rangle =
\Lambda^3 \delta^j_a, ~~~~~a,i,j=1,...,N \; ,
\label{cond}
\ee
which is in the fundamental representation of the gauge group.
In this case, one expects that there is no phase boundary between
confining and Higgs phases \cite{complementarity}. Thus, as far as symmetry
realization and
massless composites are concerned, one may as well study the Higgs phase
and consider the
effective Higgs field $\Phi^j_a$ as fundamental. In the Higgs
picture the v.e.v. (\ref{cond}) breaks both the gauge group
and the global $SU(N+4)$ symmetry of the $F$'s. However, a certain global
subgroup
$SU(N)\times SU(4)$ remains unbroken. Here $SU(N)$ is the diagonal part
of $SU_{gauge}(N)$ and the $SU(N)$ subgroup of global $SU(N+4)$,
operating
on the first $N$ of the $F$ fermions, while $SU(4)$ is the subgroup of
$SU(N+4)$ operating on the last four $F$'s which do not
participate in the condensation. In addition, there is also a global
$U(1)$ symmetry which survives the formation of the condensate
(\ref{cond}) and will be discussed further below.
In the Higgs phase, the only possible fermionic mass term
can originate from a coupling
of the form $\Phi^j_a {\bar F}^{ia} {\bar S}_{ij}$ + h.c. Upon
diagonalization,  this interaction leaves massless two sets of fermions,
transforming according to $SU(N)\times SU(4)$
as $((N\times N)_{asym},1)$ and $(N,4)$, respectively.
In terms of the original preons $F$ and $S$ these fermions may be
constructed schematically as
\be
f_{[ab]}=F_{[a}F_{b]}S; ~~~f'_{aA}=F_a F_A S,
{}~~~~~a,b=1,...,N; ~~~A=1,...,4 \; .
\label{composite}
\ee
One can check that these fermions
satisfy 'tHooft's anomaly matching conditions \cite{tHooft} for all
anomalies of the unbroken global group $SU(N)\times SU(4)\times U(1)$.

At the lagrangian level, there are two particle number $U(1)$
symmetries associated with the numbers of $F$'s and $S$'s,
respectively. However, only a certain linear combination of
these two is anomaly-free with respect to the $SU(N)$ gauge
fields.
The anomaly-free fermion number is $q=n_S (N+4)/N - n_F (N+2)/N$,
where $n_S$ and $n_F$ are the $S$ and $F$ particle numbers.
A linear combination of this anomaly-free generator
and a diagonal generator of the global $SU(N+4)$,
\be
I_{N+4}=N^{-1}~diag(1,...,1,-N/4,...,-N/4) \; ,
\label{gen}
\ee
gives the generator $q'=q-2I_{N+4}$ of the $U(1)$ symmetry that
is unbroken by the condensate (\ref{cond}) and acts on the composite
states. Indeed, the $q'$ charges of the preons:
$F_{ia}~(q'=-(N+4)/N);~F_{iA}~(q'=-(N+4)/(2N));~S^{ij}~(q'=(N+4)/N)$
guarantee that $\Phi^j_a$ has $q'=0$.

It is important to
understand how the $U(1)$ anomaly of the preon theory
manifests itself at the level of composites.
Due to complementarity, we may again analyze the theory in
the Higgs phase where the relevant fluctuations are instantons.
At very short distances,
much shorter than $\Lambda^{-1}$, where $\Lambda$ is
the dynamical scale of the theory, the Higgs v.e.v.
is inoperative, and instantons of $SU_{gauge}(N)$ give rise
to a 'tHooft effective interaction \cite{tHooft2} involving
$N+4$ $F$'s and $N+2$ $S$'s, corresponding to the numbers
of zero modes of these representations. As we go to
longer distances, the v.e.v. turns on and the zero modes coupled
to the condensate (\ref{cond}) are lifted. These correspond to
the first $N$ $F$'s and $N$ linear combinations of $S$'s.
The remaining six zero modes become those of composites, four
of them corresponding to states $f'$ and two to ${\bar f}$.
That is, the 'tHooft interaction at the preon level leads to
an effective interaction at the bound state level of the form
\be
\epsilon^{ABCD} f'_{aA} f'_{bB} f'_{cC} f'_{dD} {\bar f}^{[ab]}
{\bar f}^{[cd]} \; ,
\label{vertex}
\ee
where for notational simplicity we have suppressed Lorentz indices.
This dimension nine effective vertex is accompanied by a coupling of
order $1/\Lambda^5$
in the effective lagrangian. Thus, in chiral gauge
theories, at least where complementarity is applicable, the $U(1)$
anomaly is reflected in the low-energy effective lagrangian
by a multi-leg fermionic vertex suppressed by a high power of the
compositeness scale.

In the model described above, the composite fermions
(\ref{composite}) are strictly massless and the global chiral
symmetry $SU(N)\times SU(4)\times U(1)$ is exact.
A possible way to break this group, so that some fermions receive
small masses is to gauge an appropriate part of it \cite{Peccei1}.
At the preonic level, this corresponds to assigning to some preons
gauge charges of an additional gauge group.
A useful way to proceed is to let the
four preons, which we have denoted as $F_A$, $A=1,...,4$, form two
doublets with respect to a vector-like $SU(2)$ gauge group
characterized by a confining scale $\Lambda'\ll \Lambda$,
\be
F_A=\{F_{1\alpha}, F_{2\alpha}\}, ~~~~~\alpha=1,2 \; .
\label{doublets}
\ee
Since $\Lambda'\ll \Lambda$, the effect of the new gauge interaction
is best understood at the level of composite states.
The states labelled by $f$'s, which were $SU(4)$ singlets,
are now $SU(2)$ singlets, while the states denoted by $f'$ decompose
analogously to \eq{doublets}, $f'_{aA}=\{f'_{1a\alpha}, f'_{2a\alpha}\}$.
Because the $SU(2)$ gauge theory is vector-like, it produces
chirality-breaking condensates similar to those of QCD. Assuming
the pattern of condensation\footnote{If the condensates
$\langle \epsilon^{\alpha\beta} f'_{1a\alpha} f'_{1b\beta} \rangle$,
$\langle \epsilon^{\alpha\beta} f'_{2a\alpha} f'_{2b\beta} \rangle$
form too, the $SU(N)$ symmetry is broken to a symplectic group which
also allows mass terms for the states $f$.}
\be
\langle \epsilon^{\alpha\beta} f'_{1a\alpha} f'_{2b\beta} \rangle
= \Lambda'^3 \delta_{ab} \; ,
\label{breaking}
\ee
the global $SU(N)$ symmetry acting on composite states
is broken spontaneously to the orthogonal group $O(N)$.
Moreover, the low-energy $U(1)$ symmetry,
besides being broken by (\ref{breaking}),
is now also broken explicitly
by the anomaly associated with the $SU(2)$ gauge fields. The $O(N)$
symmetry is thus the only symmetry remaining which acts on the $f$
states, and it allows for a mass term $m_f f_{[ab]} f_{[ab]}$.
Since the $f$'s do not have $SU(2)$ quantum numbers themselves, the
only way by which the symmetry-breaking pattern is communicated to
them is through the contact interactions with
the $f'$ states, suppressed by the compositeness
scale. In particular, the anomalous interaction of \eq{vertex} is
necessary because all non-anomalous interactions preserve a
separate particle number of the $f$'s. The condensates of \eq{breaking},
in conjunction with the anomalous vertex (\ref{vertex}), give precisely
a mass term for the $f$'s with
\be
m_f \sim \frac{\Lambda'^6}{\Lambda^5} \; .
\label{mass}
\ee
Due to the difference between mass scales $\Lambda$ and $\Lambda'$,
the mass (\ref{mass}) can be made arbitrarily small compared to
either of these scales.

An unfortunate feature of this particular model, which needs
to be avoided in realistic model building, is that it typically
produces light fermions in real representations of orthogonal
or symplectic groups, leaving no room for
the weak interaction group $SU_W(2)$. In other words, since
the observed quarks and leptons
are chiral with respect to weak interactions, care
should be taken that this possibility remains open for the
bound states produced by the preon theory.
This suggests that one wants condensates, analogous to those of
\eq{breaking}, not to give masses to some of the bound state
fermions, but only to give rise to appropriate $SU_W(2)$
conserving four-fermion effective interactions tying left- and
right-handed fermions together. Masses could then be generated by another
set of condensates (technicolor), with these effective interactions
playing the role of ETC interactions. We will see how this works in a
model of one generation of quarks in the next section.

Even though the models discussed in the sequel are in many respects
different from the above simple model, some features of it will
remain relevant.
In particular, the multileg interactions produced by the instantons
of the preonic theory will be an important ingredient in our attempts
to generate quark mixing in Sect.4.
Besides, one can think of a physical context where the
mass pattern produced by our toy model may be sufficient,
namely giving Majorana masses
to right-handed neutrinos. Since the generation of such masses
does not involve $SU_W(2)$ breaking, which has a relatively low energy
scale, such Majorana masses can be arbitrarily large, precisely as is
needed to make the observable neutrinos naturally light.

\section{A one generation model of quarks}
More realistic composite models for quarks and leptons can be
constructed by making use of multiple repetitions of the preonic
model discussed in the last section. These models are not economical
in their structure, but they do provide a very nice theoretical
laboratory to test ideas. Furthermore, we note that the
complicated structures which are introduced not only produce
the "quasi-elementary" fermions wanted, but also all the necessary
symmetry breaking dynamics to generate their masses.
In the simplest version of these models, quarks and leptons are
made by different preon theories. Thus, for illustrative purposes
it will suffice to consider, to begin with, just a model for one
generation of quarks. In doing so, we will not run into a
problem with the hypercharge anomaly, because in our model the
hypercharge anomaly of quarks is cancelled by that of
techniquarks.\footnote{
Leptons are needed, though, to cancel the
global \cite{Witten} $SU_W(2)$  anomaly - that is to make the total
number of $SU_W(2)$ doublets even. With quarks alone, the number of
doublets at preonic level in the one-generation model which will
be considered is $6\times 6 + 21 = 57$.}

The model to be considered is constructed as follows.
A doublet of left-handed quarks and each of the two right-handed
quarks descend from their own $SU(N)$ preonic gauge theories, which
here we choose to have $N=6$.
Each of these preon theories produces massless composite
fermions in the (15,1) and (6,4) representations of their
respective global $SU(6)\times SU(4)\times U(1)$ groups.
Because now we have both left- and right-handed particles at
our disposal, we can gauge a common vectorial $SU(4)$,
which we will call metacolor,
as a whole, rather than gauging only an $SU(2)$ subgroup of it
as we did with the toy model of the last section.
This $SU(4)$ gauge interaction does not produce masses directly but
gives rise to four-fermion interactions between left-
and right-handed composites. Furthermore, out of a common vectorial
$SU(6)$, two SU(3) subgroups are gauged, one becoming color and the
other acting as technicolor. Finally,
the $SU_W(2)$ gauge group is built into the model by having a
doubled fermionic content
in the left-handed preon theory. Within this structure, as we shall
discuss, appropriate anomaly-free hypercharge assignments can be made.

Although this model certainly looks like an ugly mechanical aggregate,
it has only two parameters more than the standard model.
The model has 3 preonic dynamical scales, $\Lambda_L$, $\Lambda^u_R$
and $\Lambda^d_R$, and one additional
scale parameter $\Lambda_4$ for the common $SU(4)$ gauge group.
The technicolor dynamical scale $\Lambda_{TC}$ is not another
parameter, since it is related, as usual, to
the W and Z masses. These 4 parameters replace
the two Yukawa couplings of a one generation standard model of
quarks. We will see, however, that even though there are more
parameters, one gains a more dynamical understanding of how mass
differences between up- and down-type quarks can arise.

Let us now describe the one-generation model in more detail.
As we said above, the model includes three chiral gauge theories, each
based on a separate preonic gauge group $SU(6)$. One of this groups
has a doubled fermionic content, that is $2\times 10=20$
left-handed fermions in the fundamental representation and two
fermions in the conjugate symmetric representation (of dimension 21).
This doubling is intended to allow the introduction of the
$SU_W(2)$ symmetry.
At the preonic level, six out of the ten
pairs of preons belonging to fundamental representation
and the pair in the symmetric representation form $SU_W(2)$ doublets,
while the remaining $2\times 4=8$ preons in the fundamental representation
are taken as $SU_W(2)$ singlets.
With these assignments, the order parameter of the complementarity
picture $\langle F^{ai} S_{ij} \rangle\propto\delta^a_j, ~a,i,j=1,...,6$
can be made an $SU_W(2)$ singlet, so that $SU_W(2)$ is not broken at this
level. Among the six doublets $F^{ai}, ~a=1,...,6$ three
form a conjugate triplet under the color group $SU_c(3)$,
while three others
form a conjugate triplet under a technicolor gauge group which is
also an $SU(3)$.
So, in total, the left-handed preonic theory has three colored preonic
fields
$C_L$, three technicolored fields $T_L$, which are in the fundamental
representation of the preonic gauge group $SU(6)$ and are doublets with
respect to $SU_W(2)$,
an $SU_W(2)$ doublet state $S_L$ in the conjugate symmetric
representation of preonic $SU(6)$, and $2\times 4$ fields
$M_L^{u}$, $M_L^{d}$ which are in the fundamental representation of the
preonic gauge group and are $SU_W(2)$ singlets. There are two other $SU(6)$
preonic theories for the right-handed fermions, one for the up and one for
the down quarks.
Each has a single copy of the basic fermionic content. So, in addition
we have preons $C_R^{u,d}$, $T_R^{u,d}$, $S_R^{u,d}$ and $M_R^{u,d}$,
where again $C$'s denote conjugate triplets of color and $T$'s denote
conjugate triplets
of technicolor. All right-handed preons are $SU_W(2)$ singlets.

When we apply complementarity to the left-handed theory, neglecting
for the moment the color and technicolor gauge couplings, we are left
with composite states which transform as (15,1) and (6,4) under
$SU_{diag}(6)\times SU(4)\times U(1)$.
Due to the original doubling, each of
these states now comes in two varieties, but while the two copies of
(15,1) form an $SU_W(2)$ doublet, which we denote as $f_L$,
the two (6,4) states, which we denote as ${f'}^u_L$, ${f'}^d_L$
are $SU_W(2)$ singlets.
The (6,4) states are the only states which have in them the preons
$M_L^u$, $M_L^d$.
Analogously, the right handed theories produces composite
fermions $f_R^{u,d}$ and ${f'}_R^{u,d}$, all singlets under $SU_W(2)$.
If there were no color and technicolor
couplings, the states of each theory would transform under their own
separate $SU_{diag}(6)$. However, when these couplings are turned on
this $[SU_{diag}(6)]^3$ symmetry is broken explicitly to
$SU_c(3)\times SU_{TC}(3)$ (times a certain number of $U(1)$'s which
we will discuss further below).
In addition, we are now going to gauge an
$SU(4)$ in such a way that all $M$ preons, both left and right, fall
in its fundamental representation.
This breaks the global $[SU(4)]^3$ symmetry down to $SU_{gauge}(4)$.

Besides having anomalous vertices analogous to \eq{vertex},
each of the three preonic theories produces contact interactions
between the respective composites of the form
\be
({\bar f_L} \gamma_{\mu} \lambda^A f_L)
({\bar {f'}^{u,d}_L} \gamma_{\mu} \lambda^A {f'}^{u,d}_L),
{}~~~({\bar f_R^u} \gamma_{\mu} \lambda^A f_R^u)
({\bar {f'}_R^u} \gamma_{\mu} \lambda^A {f'}_R^u),
{}~~~({\bar f_R^d} \gamma_{\mu} \lambda^A f_R^d)
({\bar {f'}_R^d} \gamma_{\mu} \lambda^A {f'}_R^d) \; ,
\label{contact}
\ee
where the $\lambda^A$ are global $SU(6)$ generators.
The interactions in \eq{contact} preserve all partial fermionic
numbers, that is the fermionic numbers of each
of the composite states $f$ and $f'$ for the left and for both the
right theories separately. Through the $SU_{gauge}(4)$ interactions,
the left- and right-handed sectors finally meet.
Because the $SU_{gauge}(4)$ theory
is vector-like, one expects that vacuum condensates
\be
\langle {\bar {f'}}^u_L {f'}^u_R \rangle =
\langle {\bar {f'}}^d_L {f'}^d_R \rangle
\label{condensates}
\ee
form, and the contact interactions of \eq{contact} will give rise to
ETC interactions, as sketched in fig.1:
\be
({\bar f_L} \gamma_{\mu} \lambda^A f_L)
({\bar f_R^{u,d}} \gamma_{\mu} \lambda^A f_R^{u,d}) \; .
\label{etc}
\ee
Note that since both left- and right-handed $f'$ states are $SU_W(2)$
singlets, the condensates (\ref{condensates}) {\em do not break}
$SU_W(2)$. As we shall show below, they also preserve hypercharge.

After gauging the metacolor $SU(4)$ and color and technicolor $SU(3)$'s
in the manner indicated above, one can identify 7 chiral $U(1)$'s
which are preserved in the binding of the preonic $SU(6)$ theories.
Three of these $U(1)$'s correspond to the charges $q'$ of each preon
theory, which we had identified earlier. With a convenient rescaling,
the nontrivial preon assignments of these charges are
\[
q'_L:~~~~~\{C_L=1;~~T_L=1;~~M_L^u=1/2;~~M_L^d=1/2;~~S_L=-1\} \; ;
\]
\[
{q'}^u_R:~~~~~\{C^u_R=1;~~T^u_R=1;~~M_R^u=1/2;~~S_R^u=-1\} \; ;
\]
\be
{q'}^d_R:~~~~~\{C^d_R=1;~~T^d_R=1;~~M_R^d=1/2;~~S_R^d=-1\} \; .
\label{charges}
\ee
In addition, there are $U(1)$'s in each of the three theories,
also free from preonic anomalies,
which exploit the fact that the preons belonging to fundamental
representations now come in different types. This freedom allows the
introduction of 4 more conserved charges, with the nontrivial
preon assignments being as follows
\[
q^{u}_R:~~~~~\{C^u_R=1;~~T^u_R=-1;~~M_R^u=0;~~S_R^u=0\} \; ;
\]
\[
q^{d}_R:~~~~~\{C^d_R=1;~~T^d_R=-1;~~M_R^d=0;~~S_R^d=0\} \; ;
\]
\[
q_L:~~~~~\{C_L=1;~~T_L=-1;~~M_L^u=0;~~M_L^d=0;~~S_L=0\} \; ;
\]
\be
{\bar q}_L:~~~~~\{C_L=1;~~T_L=-1;~~M_L^u=1/2;~~M_L^d=-1/2;~~S_L=0\} \; .
\label{morecharges}
\ee
Only 4 combinations of the above chiral $U(1)$'s
do not have any metacolor, technicolor or color anomalies. Two
of these $U(1)$'s can be chosen in a manifestly vector-like fashion
already at the preon level, namely
those corresponding to the charges
\[
q'_V=q'_L + {q'}_R^u + {q'}_R^d \; ;
\]
\be
q_V=q_L + q_R^{u} + q_R^{d} \; ,
\label{vectorcharges}
\ee
while the other two $U(1)$'s are still chiral.
The charges for these latter $U(1)$'s can be taken as
\[
{\tilde q}_L={\bar q}_L - q_L \; ;
\]
\be
{\tilde q}_R={q'}_R^u - {q'}_R^d \; .
\label{chiralcharges}
\ee
We note that three of these $U(1)$'s also have no $SU_W(2)$ anomaly,
but the fourth one, associated with $q'_V$, has such an anomaly.

The charges (\ref{morecharges}) are not exactly the ones
preserved in the preonic binding. One can see this by noticing
that the order parameter (\ref{cond}) of the complementarity
picture is not neutral with respect to these charges. However,
the order parameter is neutral with respect to certain linear
combinations of these charges and diagonal generators of the
corresponding preonic $SU(6)$ gauge groups,
\[
Q_R^u=q_R^u + diag(1,1,1,-1,-1,-1)^u_R \; ;
{}~~~~Q_R^d=q_R^d + diag(1,1,1,-1,-1,-1)^d_R \; ;
\]
\be
Q_L=q_L + diag(1,1,1,-1,-1,-1)_L \; ;
{}~~~~{\bar Q}_L={\bar q}_L + diag(1,1,1,-1,-1,-1)_L \; .
\label{comb}
\ee
The vector-like charge $q_V$ of \eq{vectorcharges}
is modified accordingly,
\be
Q_V=Q_L+Q_R^u+Q_R^d \; .
\label{vectorcomb}
\ee
At the level of composite states, there is no difference between
$q_R^u$ and $Q_R^u$, etc. because the composite states are neutral
with respect to the preonic gauge groups.

The metacolor condensates of \eq{condensates} obviously
preserve the two vector-like charges $q'_V$ and $Q_V$. However,
they break ${\tilde q}_L$ and ${\tilde q}_R$ individually,
preserving a linear combination of them. Since the $f'$ states
have the following ${\tilde q}_L$ and ${\tilde q}_R$ assignments
\[
{\tilde q}_L:~~~~~\{{f'}^u_L=1/2;~~{f'}^d_L=-1/2;~~{f'}^u_R=0;
{}~~{f'}^d_R=0\} \; ;
\]
\be
{\tilde q}_R:~~~~~\{{f'}^u_L=0;~~{f'}^d_L=0;~~{f'}^u_R=1/2;
{}~~{f'}^d_R=-1/2\} \; ,
\label{fprimecharges}
\ee
what is preserved by (\ref{condensates}) is
\be
{\tilde q}_V={\tilde q}_R + {\tilde q}_L \; .
\label{preserved}
\ee
A linear combination of $Q_V$ and ${\tilde q}_V$,
\be
Y=\frac{1}{6} Q_V + {\tilde q}_V \; ,
\label{Z}
\ee
may be gauged without acquiring an anomaly and is identified
with the hypercharge. Therefore, the two anomaly-free charges
surviving down to the technicolor scale may be taken as $Q_V$ (or,
at that scale, $q_V$) and the hypercharge.

With respect to the gauged $SU_c(3)\times SU_{TC}(3)$ group,
the 15 composite states $f$
transform as $(3,1)+({\bar 3},{\bar 3})+(1,3)$. The first component
corresponds to the observable quarks, while the remaining two are
techniquarks. The gauge group $SU_{TC}(3)$ causes techniquarks
to condense, thus breaking $SU_W(2)$ and giving masses to the quarks.
The hypercharge symmetry is also broken at this stage, while the
vectorial symmetry associated with $q_V$ is preserved.
Let us estimate the dependence of quark masses on the various
mass scales present in the theory.
The contact interactions \eq{contact} are multiplied by
factors of $\Lambda_{preon}^{-2}$ in the effective lagrangian,
$\Lambda_{preon}$ being the scale of the corresponding
preonic theory,
$\Lambda_{preon}=\{\Lambda_L, \Lambda_R^u, \Lambda_R^d\}$.
Hence, the ETC interactions of \eq{etc} and fig.1 are multiplied by
factors
\be
\frac{1}{\Lambda_{ETC}^{u~2}}=
\frac{\Lambda_4^2}{\Lambda_L^{2} \Lambda_R^{u~2}} \; ;
{}~~~\frac{1}{\Lambda_{ETC}^{d~2}}=
\frac{\Lambda_4^2}{\Lambda_L^{2} \Lambda_R^{d~2}} \; ,
\label{lambda}
\ee
where $\Lambda_4$ is the scale of $SU_{gauge}(4)$ theory.
The order of magnitude of quark masses is given
by $\Lambda_{TC}^3/\Lambda_{ETC}^2$. In virtue of \eq{lambda}
this gives
\be
m_u\simeq
\frac{\Lambda_{TC}^3 \Lambda_4^2}{\Lambda_L^{2}
\Lambda_R^{u~2}} \; ;
{}~~~m_d\simeq
\frac{\Lambda_{TC}^3 \Lambda_4^2}{\Lambda_L^{2}
\Lambda_R^{d~2}} \; .
\label{tcmass}
\ee
The characteristic feature of this type of models,
therefore, is that the
scale of compositeness is in inverse relation to the mass: the heavier
the particle, the larger is its 'size'.
If we are to apply formulas like (\ref{tcmass}) to reality,
the lowest compositeness scale is that for the top quark.
The highest possible value for this scale is achieved if we assume
$\Lambda_4\sim\Lambda^{(3)}_L\sim\Lambda^t_R$, where $\Lambda^{(3)}_L$
is the scale of the left-handed preonic theory for the third generation.
Then, for $m_t\sim 100$ GeV, one gets
\be
\Lambda_4\sim\Lambda_L^{(3)}\sim\Lambda^t_R\sim\mbox{3~TeV} \; .
\label{top}
\ee
These estimates are too naive and should
be modified in models where the technicolor
coupling runs sufficiently slowly (walking technicolor \cite{walking})
or has a fixed point \cite{nonfree}.
Since the above estimate serves mostly for illustrative purposes,
we do not discuss these further refinements here.

The existence of an additional scale
$\Lambda_4$, which in general is intermediate between the
technicolor scale
(1 TeV) and the scales of compositeness of light particles,
is very welcome from a phenomenological point
of view. In our model there are no global symmetries preserved by chiral
binding and $SU_{gauge}(4)$ and broken by technicolor.
Hence, $\Lambda_4$ is the scale of 'vacuum alignment', where
the gigantic global symmetry of the preonic theory is broken by the
condensates
(\ref{condensates}), in addition to being broken by the gauge couplings.
If $\Lambda_4$ can be made sufficiently large,
the resulting pseudoGoldstone bosons will be harmless phenomenologically.
In the absence of such an intermediate scale, the global symmetry
would be broken
by technicolor condensates resulting in lower masses for the
pseudoGoldstones.
In addition to relatively heavy pseudoGoldstones, there are
several strictly massless  particles in our model, associated with
exact anomaly-free global $U(1)$ symmetries broken spontaneously
at various levels. Again, none of these symmetries are broken
exclusively by technicolor, so the scales of derivative couplings
of Goldstone particles
to ordinary matter are at least of order $\Lambda_4$, which
in principle allows to put them beyond experimental detection.

\section{Prospects for quark mixing}
The simplest way to accommodate three generations of quarks in our
model is to enlarge the model in a mechanical way, so that
each left-handed quark doublet and each right-handed
singlet is prepared by its own theory and has its own compositeness
scale. This increases the number of preonic theories to nine.
All preons, however, share the same
$SU_{gauge}(4)$, technicolor, color, $SU_W(2)$ and  hypercharge
interactions.
The masses for all quarks can be generated by the same mechanism as
before, but problems arise when one tries to obtain quark mixing.
These problem are related to the existence of certain vectorial
symmetries, one per generation, which protect the quarks from mixing.
At the simplest level, there are restrictions imposed by the existence
of three conserved charges, analogous to $q_V$ of the previous section.
Although this is not exactly where the worst part of the problem comes
from, it is nevertheless a good prototypical example to begin our
discussion.
Because mass generation in our model is due, essentially, to
vector-like gauge interactions, these vectorial symmetries might be
expected to be preserved in binding and thus to prohibit mixing.
Therefore, to understand how quarks mixing can arise in this type of
models, we have to consider possible deviations of the $SU_{gauge}(4)$
and technicolor interactions from vector-like behavior.

A possible source of such deviations are the $SU_W(2)$ interactions
which, in the presence of multiple fermionic species, may grow
strong at the $SU_{gauge}(4)$ dynamical scale. The $SU_W(2)$ interactions
are certainly not vector-like because they involve only left-handed
particles. However, an additional agent is needed to communicate this
information to the states $f'$, which are
the only ones with $SU_{gauge}(4)$ quantum numbers,
because all these states, both left and right, are $SU_W(2)$ singlets!
The only such agent are the preonic interactions, and for these to be
effective, the $SU_{gauge}(4)$ scale should be close
to that of at least one of the preonic theories, say, that of the
right-handed top-quark. If some of the preonic interactions are
not completely screened at the scale of $SU_{gauge}(4)$, then
these interactions themselves can be a source of non-vector-like
behavior. In these circumstances, it is conceivable that non-diagonal,
flavor mixing condensates of the states $f'$ can be formed
in the $SU_{gauge}(4)$ binding, for example
\be
\langle {\bar {f'}}_L^d {f'}_R^s \rangle \neq 0 \; .
\label{nondiag}
\ee
If (\ref{nondiag}) obtains, then the vectorial $U_V(1)$ symmetries are
broken dynamically. However, even in this case one still has further
difficulties.

The real problem with quark mixing arises when one tries to
communicate the breakdown manifested by \eq{nondiag} to the observable
fermions which reside in the multiplets $f$. One notices that
while the ETC contact interactions between $f$ states of different
generations are now possible, they will always be of the form
\footnote{Henceforth, the $f_L$'s and their compositeness scales will
carry a generation superscript.},
\be
({\bar f^{(1)}_L} \gamma_{\mu} \lambda^A f^{(1)}_L)
({\bar f_R^{s}} \gamma_{\mu} \lambda^A f_R^{s}) \; ~~~~\mbox{etc.}
\label{uds}
\ee
That is, while $f$'s from different generations now can interact,
their flavor numbers are still conserved. There is a symmetry
reason for this behavior. Indeed, as in the toy model of the previous
section, unless the anomalous interactions generated by instantons of
the preonic theories are included, the fermionic numbers of
the $f$ and $f'$ states are conserved separately.
As long as these interactions are neglected, the separate flavor
symmetries of the states $f$ will keep the observable fermions
from mixing, even in the presence of non-diagonal condensates of the
states $f'$, \eq{nondiag}. Therefore, to get mixing, one needs
to consider both ETC interactions (\ref{uds}) {\em and} the
instanton-induced vertices. Doing so, however, one if immediately
faced with two problems.
First, the instanton-induced vertices appear to be too small
in magnitude to account for the observable mixing. Second - and this
is a more profound problem - though the preonic instantons do
break the separate fermionic numbers of the $f$'s, they preserve
the $Z_2$ subgroups of the corresponding $U(1)$ symmetries! This
follows simply from the fact that, as in the toy model of Sect.2,
each instanton has two zero modes of the states $f$ of the corresponding
flavor associated with it. Moreover, it is difficult to imagine
that vector-like combinations of these $Z_2$ symmetries
can be broken spontaneously by technicolor, since the non-vector-like
interactions are already weak at the technicolor scale, so they cannot
drive the system away from vector-like behavior.
Thus, these $Z_2$ symmetries should survive to low energies
and are sufficient to prohibit the mixing of quarks. Because of
this, the simplest family generalization of our one-generation
model is unrealistic. Although there are other ways to introduce families
in these models, we shall not discuss them further here.

It seems, however, not entirely out of place to discuss for
the remainder of this section some
general features of mixing matrices expected in this type
of composite models, assuming that eventually a more realistic model
of this type can be constructed. By "this type of models" here
we mean models where, as in the one-generation model of
Sect.3, the masses of quarks are in inverse relation to their
compositeness scales. If a non-trivial mixing can be generated,
we presume that this property will hold both for the diagonal and
non-diagonal entries of the two mass matrices - for the up- and
down-type quarks.

The CKM matrix appears after diagonalization of the
two quark mass matrices.
There is no reason why these matrices should be symmetric, because
the left and right components of the quarks have different
substructure. This means that, for example, the mass matrix for up-type
quarks
\be
{\hat M}_{RL}= \left(
\begin{array}{lll}
 M_{uu} & M_{uc} & M_{ut} \\
 M_{cu} & M_{cc} & M_{ct} \\
 M_{tu} & M_{tc} & M_{tt}
\end{array}\right)
\label{M}
\ee
should be diagonalized with the help of two unitary $3\times 3$
mixing matrices: one, denoted by ${\hat O}_R$, acting on right-handed
quarks, another, ${\hat K}_u$, on the left-handed quarks, so that
\be
{\hat O}_R {\hat M}_{RL} {\hat K}_u^{\dagger}=
diag(m_u, m_c, m_t)\equiv {\hat m}_u \; .
\label{diag}
\ee
In the standard model, it is a matter of choice whether
to associate mixing with up- or down-type quarks, or some
linear combination,  because the
only observable effect is the product of the mixing
matrices for left-handed up- and down-type quarks - the CKM matrix
$V=K^{\dagger}_d K_u$.
The mixing matrices for right-handed quarks, e.g. ${\hat O}_R$,
have no observable effect in the standard model.

In the kind of models under discussion, however, in addition to all
the standard interactions, there are four-fermion interactions
of the type of eqs.(\ref{etc}),(\ref{uds}).
When two of the legs in \eq{etc} are chosen to be quarks
and other two techniquarks, these terms play the role of ETC
interactions. When all four legs are quarks, they are the
new interactions between observable particles and a potential
source of FCNC. Unlike the standard model interactions, these
new terms are affected by up- and down-, and left- and
right-mixings separately. It is easy to imagine now that due
to the strong dependence of the non-diagonal entries of the mass
matrices on the corresponding compositeness scales and, hence, on
the particle masses, the largest mixing occurs for the heaviest
particles (we will be a bit more precise below).
That is, the CKM matrix should be dominated
by mixing of the up-type quarks and, hence, the up-type sector
is where the FCNC effects will be the biggest. Thus,
composite models of the type described here suggest
that the primary place to search experimentally for FCNC effects is
in the $D_1-D_2$ mass difference. Coincidentally, FCNC
effects in the up-type sector (e.g. in $D$-mesons) are less thoroughly
studied experimentally than those in the down-type system.

If the mixing of down-type quarks is neglected, the matrix
${\hat K}_u$ is precisely the observable CKM matrix $V$. In this case,
one can obtain the up-type quark mass matrix ${\hat M}_{RL}$
in terms of the parameters of the standard model by considering the matrix
equation following from \eq{diag},
\be
{\hat M}_{RL}^{\dagger} {\hat M}_{RL} =
{\hat K}_u^{\dagger} {\hat m}^2 {\hat K}_u \; .
\label{squared}
\ee
The matrix equation (\ref{squared})
constitutes nine real equations for the nine complex entries of the
matrix ${\hat M}_{RL}$. So, in general this matrix is only
half-determined
by \eq{squared} (if we knew the matrix ${\hat O_R}$, another
half of the equations would come from there).
Combined with our model considerations, however, \eq{squared} is
sufficient, essentially because we anticipate a certain
hierarchy among the elements of ${\hat M}_{RL}$.
First (and unrelated to our model considerations), we use the
arbitrariness of phases of the right-handed quarks
to make $M_{uu}$, $M_{cc}$ and $M_{tt}$ real and positive.
Second, because the terms $M_{uc}$ and $M_{cu}$ involve the two lighter
(most elementary) quarks, they should be
the smallest non-diagonal entries within our model, so we drop them.
Finally, the term
$M_{ut}$ appears in ${\hat M}_{RL}^{\dagger} {\hat M}_{RL}$ in the
combinations $M_{uu}^* M_{ut} + M_{tu}^* M_{tt}$ and
$|M_{ut}|^2+ M_{tt}^2$, so given that $M_{ut}$ is of the same order
as $M_{tu}$ or smaller, its contributions to ${\hat M}^{\dagger} {\hat M}$
are negligible. So $M_{ut}$ also drops out and we are left
with nine equations for nine real parameters.

Among the parameters of the standard model, the most uncertain
are the mass of the top-quark, for which we allow the range from
100 to 300 GeV, and the CP-breaking phase $\delta$, for
which we allow the range from 0 to $\pi/2$.
For the two other up-quark masses we use $m_u=5$ MeV and $m_c=1.5$ GeV.
For the CKM mixing angles we use the central values suggested in
ref. \cite{Peccei3}: $\sin\theta_1=0.22$,
$\sin\theta_2=0.95 \sin^2\theta_1$,
$\sin\theta_3=0.64 \sin\theta_1 \sin\theta_2$.
With these values, it turns out that $M_{uu}$ is
essentially independent of the choice of $m_t$ and $\delta$,
$M_{uu}=5$ MeV. Also, with good accuracy $M_{tt}\approx m_t$,
almost independently of the value of $\delta$. The absolute values of the
other matrix elements are plotted as functions of $m_t$ and $\delta$
in fig.2. These values are consistent with the hierarchy
suggested by our model.

\section{Concluding remarks}
Because our discussion has ranged over both technical issues as well as
some rather phenomenological points, it is worthwhile to try to summarize
the principal results obtained. The central result of this paper is that
it is possible to obtain very light fermions in confining gauge theories
($m_f \ll \Lambda_{comp}$) by appropriately gauging in a vector-like
manner a subset of preons in the theory. Although this result
is interesting per se, for this mechanism to be relevant for composite
models of quarks and leptons, one must envisage that mass generation occurs
through two distinct stages. In the first stage, as a result of the gauging
of a vector-like subset of preons, one establishes interactions between
the right- and left-handed components of the massless fermionic bound
states of the underlying preonic theory. These effective interactions then
give rise to masses for the fermions, when a further vector-like technicolor
interaction is switched on.

In the text a model for one generation of quarks is developed along these
lines. Interestingly, the model exhibits both a hierarchy between the obtained
fermion masses and the natural scale of the binding in the theory, as well
as a hierarchy between the up and down quark masses. Unfortunately, it
appears to be difficult to incorporate a family structure along these lines.
Although fermions can ensue as simple mechanical repetitions with quite
different masses for the bound state fermions, it is difficult to eliminate
all vestiges of natural flavor symmetries in the model. As a result,
even in the presence of hierarchical fermionic masses, it is not possible
to generate dynamically any CKM mixing. More precisely put, although
triggering elements exist for the spontaneous breakdown of the remaining
discrete family symmetries, the strength of these interactions seems far
too small to guarantee that such a breakdown really happens.

Even though we have not been able to construct any realistic model of
quarks and leptons along these lines, the above model considerations suggest
the possible telltale signals of this class of composite models.
The most immediate of these concerns the presence
of flavor-changing neutral currents (FCNC) in the theory. In these scenarios
it is not that FCNC never appear, but rather they should primarily affect
the heaviest of the quarks and leptons. Because FCNC in the heavy-quark
states are not well studied experimentally (if at all!), it is quite
possible that rather large violations of flavor conservation could be
associated with the top quark. As the exercise at the end of the paper
shows, it is possible to envisage hierarchical mass matrices along these
lines which give rise to the experimentally observed pattern of
masses and mixings for the quarks. Whether one can construct a dynamical
model which really produces this kind of mass matrices, however,
remains an open challenging problem.

\section{Acknowledgements}
One of us (RDP) is supported in part by the Department of Energy under
contract DE-AT03-88ER 40384 Mod A006 - Task C. The other one (SK) is
the holder of the Julian Schwinger fellowship in the Department of Physics
at UCLA. RDP is grateful for discussions on these matters with A. Casher,
J. Sonnenschein and S. Yankielowicz, as well as for the hospitality of the
Department of Physics of Tel Aviv University where some of this work was
carried out.

\newpage
\vspace*{1.5in}
\begin{center}
\large{FIGURE CAPTIONS}
\end{center}
\vspace{0.4in}
\noindent
FIG.1. Emergence of ETC interactions in the composite model.
\vspace{0.2in} \\
\noindent
FIG.2. Absolute values of the elements of the up-type quark mass
matrix as functions of the top-quark mass $m_t$ and the CP non-conserving
phase $\delta$.


\begin{thebibliography}{99}
\bibitem{Peccei2}
R. D. Peccei, preprint DESY 87-050 (1987); In: Proc. of the
1987 Lake Louise Winter Institute, Lake Louise, Alberta, Canada.
\bibitem{tHooft}
G. 'tHooft, In: Recent developments in gauge theories,
Cargese Lectures 1979 (Plenum, New York, 1980).
\bibitem{tc}
L. Susskind, Phys. Rev. {\bf D20}, 2619 (1979); S. Weinberg,
Phys. Rev. {\bf D13}, 974 (1976); {\bf D19}, 1277 (1979).
\bibitem{etc}
S. Dimopoulos and L. Susskind, Nucl. Phys. {\bf B155}, 237 (1979);
E. Eichten and K. Lane, Phys. Lett. {\bf B90}, 125 (1980).
\bibitem{early}
J. Preskill, In: Particles and fields 1981: testing the standard model,
Proc. Santa Cruz 1981 (AIP, New York, 1982) p.572;
I. Bars, Nucl. Phys. {\bf B208}, 77 (1982).
\bibitem{diseases}
S. Dimopoulos and J. Ellis, Nucl. Phys. {\bf B182}, 505 (1981);
E. Farhi and L. Susskind, Phys. Rep. {\bf 74C}, 277 (1981).
\bibitem{Georgi}
R. S. Chivukula and H. Georgi, Phys. Lett. {\bf 188B}, 99 (1987);
R. S. Chivukula, H. Georgi and L. Randall, Nucl. Phys.
{\bf B292}, 93 (1987); L. Randall, preprint MIT-CTP-2112 (1992);
H. Georgi, Harvard preprint HUTP-92/A037 (1992).
\bibitem{nonfree}
B. Holdom, Phys. Rev. {\bf 24}, 1441 (1981); Phys. Lett. {\bf B150},
301 (1985); K. Yamawaki, M. Bando and K. Matumoto, Phys. Rev. Lett.
{\bf 56}, 1335 (1986); T. Akiba and T. Yanagida, Phys. Lett. {\bf B169},
432 (1986).
\bibitem{inequalities}
D. Weingarten, Phys. Rev. Lett. {\bf 51}, 1830 (1983);
C. Vafa and E. Witten, Nucl. Phys. {\bf B234}, 173 (1984).
\bibitem{complementarity}
K. Osterwalder and E. Seiler, Ann. Phys. {\bf 110}, 440 (1978);
E. Fradkin and S. H. Shenker, Phys. Rev. {\bf D19}, 3682 (1979);
T. Banks and E. Rabinovici, Nucl. Phys. {\bf B160}, 349 (1979);
S. Dimopoulos, S. Raby and L. Susskind, Nucl. Phys. {\bf B173},
208 (1980).
\bibitem{BY} I. Bars and S. Yankielowicz, Phys. Lett. {\bf B101}, 159
(1981).
\bibitem{MAC} S. Dimopoulos, S. Raby and L. Susskind, Nucl. Phys. {\bf B169},
373 (1980).
\bibitem{tHooft2} G. 'tHooft, Phys. Rev. Lett. {\bf 37}, 8 (1976);
Phys. Rev. {\bf D14}, 3432 (1976).
\bibitem{Peccei1}
J. Bijnens, R. D. Peccei and D. Zeppenfeld, unpublished.
\bibitem{Witten}
E. Witten, Phys. Lett. {\bf 117B}, 324 (1982).
\bibitem{walking} T. Appelquist, D. Karabali and L. C. R. Wijewardhana,
Phys. Rev. Lett. {\bf 57}, 957 (1986).
\bibitem{Peccei3}
R. D. Peccei, preprint UCLA/91/TEP/40 (1991); In: Proc. of the 1991
Meeting of the Division of Particles and Fields of the American
Physical Society, Vancouver, B.C. Canada, 1991.
\end{thebibliography}
\end{document}